\documentclass[aps,prd,twocolumn,showpacs,nofootinbib,amsmath,amssymb,floatfix,superscriptaddress,showkeys]{revtex4}
\usepackage{graphicx}% Include figure files
\usepackage{dcolumn}% Align table columns on decimal point
\usepackage{bm}% bold math
\usepackage{captcont}
\usepackage{xcolor}

\usepackage{epstopdf}
\usepackage{mathtools}
\usepackage{natbib}
\usepackage{ulem}
\definecolor{blue0}{rgb}{0,0,0.6}
\usepackage[colorlinks,linkcolor=blue0,anchorcolor=blue0,citecolor=blue0,urlcolor=blue0]{hyperref}

\newcommand{\jcap}{J. Cosmol. Astropart. Phys.}
\newcommand{\apjl}{Astrophys. J. Lett.}
\newcommand{\physrep}{Phys. Rep.}
\newcommand{\mnras}{Mon. Not. R. Astron. Soc.}

\newcommand{\aap}{Astron. Astrophys}

\begin{document}
\title{Is the gamma-ray emission around the Coma galaxy cluster a dark matter origin?}
\author{Yun-Feng Liang}
\affiliation{Key Laboratory of Dark Matter and Space Astronomy, Purple Mountain Observatory, Chinese Academy of Sciences, Nanjing 210008, China}
\author{Zi-Qing Xia}
\affiliation{Key Laboratory of Dark Matter and Space Astronomy, Purple Mountain Observatory, Chinese Academy of Sciences, Nanjing 210008, China}
\affiliation{School of Astronomy and Space Science, University of Science and Technology of China, Hefei, 230026, China}
\author{Shao-Qiang Xi}
\affiliation{School of Astronomy and Space Science, Nanjing University, Nanjing 210093, China}
\affiliation{Key laboratory of Modern Astronomy and Astrophysics, Nanjing University, Ministry of Education, Nanjing 210093, China}
\author{Shang Li}
\email{lishang@pmo.ac.cn}
\affiliation{Key Laboratory of Dark Matter and Space Astronomy, Purple Mountain Observatory, Chinese Academy of Sciences, Nanjing 210008, China}
\affiliation{University of Chinese Academy of Sciences, Yuquan Road 19, Beijing, 100049, China}
\author{Zhao-Qiang Shen}
\affiliation{Key Laboratory of Dark Matter and Space Astronomy, Purple Mountain Observatory, Chinese Academy of Sciences, Nanjing 210008, China}
\affiliation{University of Chinese Academy of Sciences, Yuquan Road 19, Beijing, 100049, China}
\author{Yi-Zhong Fan}
\email{yzfan@pmo.ac.cn}
\affiliation{Key Laboratory of Dark Matter and Space Astronomy, Purple Mountain Observatory, Chinese Academy of Sciences, Nanjing 210008, China}
\affiliation{School of Astronomy and Space Science, University of Science and Technology of China, Hefei, 230026, China}

\date{\today}

\begin{abstract}
Recently, gamma-ray emission in the direction of Coma, with a TS value of $\sim 40$, has been reported. In this work we will discuss the possibility of such a residual emission coming from dark matter annihilation. Our results show that the gamma-ray emission within the Coma region is also spatially correlated to the mass distribution derived from weak gravitational lensing measurements very well. However the dark matter models are not supported by the spectral analysis results and constraints by observations of other targets. Thus we derive the upper limits of the dark matter annihilation cross section according to the observation of the Coma region.
\end{abstract}

\maketitle

\section{Introduction}

Galaxy clusters (GCls) comprised of hundreds of thousands of galaxies are the largest gravitational bounding objects in the Universe. Historically, the study towards the GCls gives the first indication of the existence of dark matter (DM), which is now supported by many other astrophysical and cosmological phenomena \cite{bertone05review}.
%The amount of the DM hosted in a GCl is about ?-? times that of the visible matters (i.e., stars and gases) according to today's observation \cite{ddd}. %I can't find a proper citation.
The amount of DM hosted in a GCl dominates that of the visible matters (i.e., stars and gases) according to today's observation. Hence GCls may also generate considerable gamma-rays by the DM particles within them if DM can annihilate or decay to standard model particles with relatively high rate. These gamma-ray signals could be detected by currently operating telescope such as Fermi-LAT \cite{atwood09LAT} and DAMPE \cite{dampe}, making GCls also ideal targets for DM indirect searches besides the dwarf spheroidal galaxies (dSphs) and the Galactic center.

The DM signal flux from astrophysical objects such as GCls is predicted to be related to the DM amount within them and the distance of the GCls to the Earth. Namely, the annihilation flux is proportional to a quantity named {\it J-factor}, which is the integration of the DM density $\rho^2(r)$ along the line of sight. Therefore, searching for the signals from some nearby and massive GCls is more promising. Such efforts include the works by \cite{ando12_fornax} for Fornax cluster, \cite{fermi15_virgo} for Virgo and so on \cite{fermi10_gclsDM,dugger10_dsphGcls,han12_gcls_constraints}.
Besides searching for DM signals from single nearby massive GCl, many works also attempted to enhance the search sensitivity by analyzing a sample of GCls.
Two major approaches were developed for such a purpose. One is using stacking/combining methods to analyze several to tens of GCls simultaneously based on the idea that DM parameters should be identical for all the GCls \cite{huang12cluster,anderson16gclsLine,liang16gclsLine,lisanti17_gcls,lisanti17_methods}; the other is to detect the cross-correlation between the diffuse gamma-ray sky and the catalogs of GCls in other wavelength \cite{branchini17crossCorr}.
Some tight constraints on the DM parameters have been obtained for both single source and large sample analyses \cite{yuan10gcls,fermi10_gclsDM,dugger10_dsphGcls,ando12_fornax,huang12cluster,han12_gcls_constraints,fermi15_virgo,anderson16gclsLine,liang16gclsLine,lisanti17_gcls,chan17}.

In view of the much lower J-factor of GCls than dSphs \cite{charles16sensitivity}, it seems unlikely to detect signals from GCls on the premise of the fact that no significant signal appears in the directions of dSphs. However, $N$-body numeric simulation in the $\Lambda$CDM cosmology have revealed that a cluster size DM halo will host large numbers of subhalos which are local concentration of the DM particles \cite{diemand08VL2,springel08Aquarius}. Since the flux of DM annihilation is proportional to $\rho^2(r)$, where $\rho$ is DM density, these concentrated subhalos are expected to enhance the DM signal significantly compared to the smooth DM distribution. Thus the real J-factor of a galaxy cluster should be $J=(1+b_{\rm sh}){J_{\rm sm}}$.
Currently, the determination of the boost factor (BF), $b_{\rm sh}$, still suffers from large uncertainties. With different consideration of the minimum mass and mass function of subhalos, the $b_{\rm sh}$ for galaxy clusters can vary from the range of $\sim10-50$ \cite{sc14cm,moline17bf} up to $\sim1000$ \cite{Gao2012bf1e3}.

Dark matter annihilation is not the only mechanism may generate extended gamma-ray emission from GCls. It is believed that a large amount of cosmic rays are accelerated by the merger or accretion shocks in the GCls, which may also produce significant gamma-rays by interacting with the intracluster medium (ICM).
The processes of the CR interaction often lead to different shape of the gamma-ray spectrum comparing to DM annihilation, thus one can distinguish between the two through the analysis of the energy spectra if the statistics is sufficient.
Since we are discussing the scenario of dark matter annihilation in this paper, we won't make too much introduction to the CR processes. We refer the readers to \cite{fermi10_gcls,brunetti12coma,zandanel14coma,Ackermann2014GC,fermi15_virgo,fermi16_coma,Reiss17_viralShock,Keshet17_viralShock} for related contents.

For most of previous efforts that searched for gamma-rays from GCls in the Fermi-LAT data, no significant gamma-ray emission was detected except that emitted by the central AGNs.
%\cite{fermi10_gcls,fermi10_gclsDM,huang12cluster,han12_gcls_constraints, Ackermann2014GC, fermi15_virgo, anderson16gclsLine, fermi16_coma}.
However, there are also some potential signal pronounced in the literatures.
Han et al. \cite{2012arXiv1201.1003H} reported that fitting the Virgo galaxy cluster with a very extended spatial template resulted in a signal of $\sim4.4\sigma$. But this putative signal was later on proved to offset from the centers of both Virgo and M49, and are more likely to come from incomplete modeling of the IEM and Loop I components \cite{han12_gcls_constraints,fermi15_virgo}.
Liang et al. has searched for the monochromatic signal from 16 nearby GCls and found a tentative excess at the energy of $\sim$43 GeV \cite{liang16gclsLine}. They suggested that if such an excess arise from dark matter annihilation, the average boost factor of these GCls should be high up to $\sim1000$. Disregarding the issue of requiring a very large BF, such a signal is theoretically interpretable \cite{feng16_43gev,2017arXiv170300685B}.
Recently, Xi et al. \cite{xi17coma} (hereafter X17) found that there exist some residual emissions spatially correlated to the radio halo of the Coma galaxy cluster with a maximal TS value of $\sim46$ corresponding to the significance of $>6\sigma$. They attributed these emissions to CR interactions within Coma. Such a high TS value of Coma has also been reported by ref. \cite{lisanti17_gcls}. In this work, we will test the possibility of such a emission coming from DM annihilation. We will revisit the analysis procedure by X17, but with some special consideration that accommodate to the signal from DM. Specifically, we will adopt some spatial and spectral models different from those used in X17 to perform our analysis.
If without a BF taken into account, the parameters that can lead to detectable DM signals from GCls have been ruled out by the state-of-the-art dSph observation \cite{fermi2015dsph6yr}, thus we concentrate on annihilation signal and won't discuss the decay DM in this work.

\section{Coma galaxy cluster and Fermi-LAT data analysis}
The Coma cluster has a virial mass of $M_{200}\simeq1.0\times{10^{15}\,M_{\odot}}$, together with its relatively low redshift of $z=0.0232$, making it among the clusters with highest $J$-factors \cite{reiprich02_106,chen07_106,ando12_fornax}. The Coma virial radius is about $2.2\,{\rm Mpc}$, corresponding to an angular extension of $1.2^\circ$ in the sky. It is located at very high galactic latitude ($l=58.09^\circ$, $b=87.96^\circ$; $\alpha_{\rm J2000}=194.95^\circ$, $\delta_{\rm J2000}=27.98^\circ$) where the contamination from the galactic diffuse gamma-ray background is very low. Unlike M87 hosted in the center of Virgo cluster \cite{fermi15_virgo}, no central AGN is found in the LAT data within the Coma. Furthermore, there is no bright gamma-ray point source within the $10^\circ$ region around the Coma center. All the facts listed above make Coma an ideal target to search for both CR and DM induced gamma-ray emission as well.

In this work, we will analyze the Fermi-LAT \cite{atwood09LAT} observation towards the Coma region in the energy range of 300 MeV to 500 GeV to study the tentative extended emission found by X17.
We use the publicly released Fermi-LAT Pass 8 data (P8R2\_SOURCE\_V6, FRONT+BACK) and the updated Fermi Science Tools package of version v10r0p5 to perform the analysis. We consider about 9 years data set from 2008 August 4 to 2017 July 26 (i.e., Mission Elapsed Time (MET) range 239557417-522806178).
For reducing the contamination from the Earth's limb, we remove the $\gamma$-ray events with the zenith angle greater than $90^{\circ}$. To ensure the spacecraft was in a good condition and the data set is valid for science use, the recommended quality-filter cuts (DATA\_QUAL==1 \&\& LAT\_CONFIG==1) are applied.
Following X17, we perform an unbinned likelihood analysis to achieve better detection sensitivity.
In consideration of the 300 MeV lower bound of the data set we use, the radius of the region of interest (ROI) adopted in the analysis is $7^\circ$.

\begin{figure}[hbt]
\begin{center}
\includegraphics[width=0.95\columnwidth]{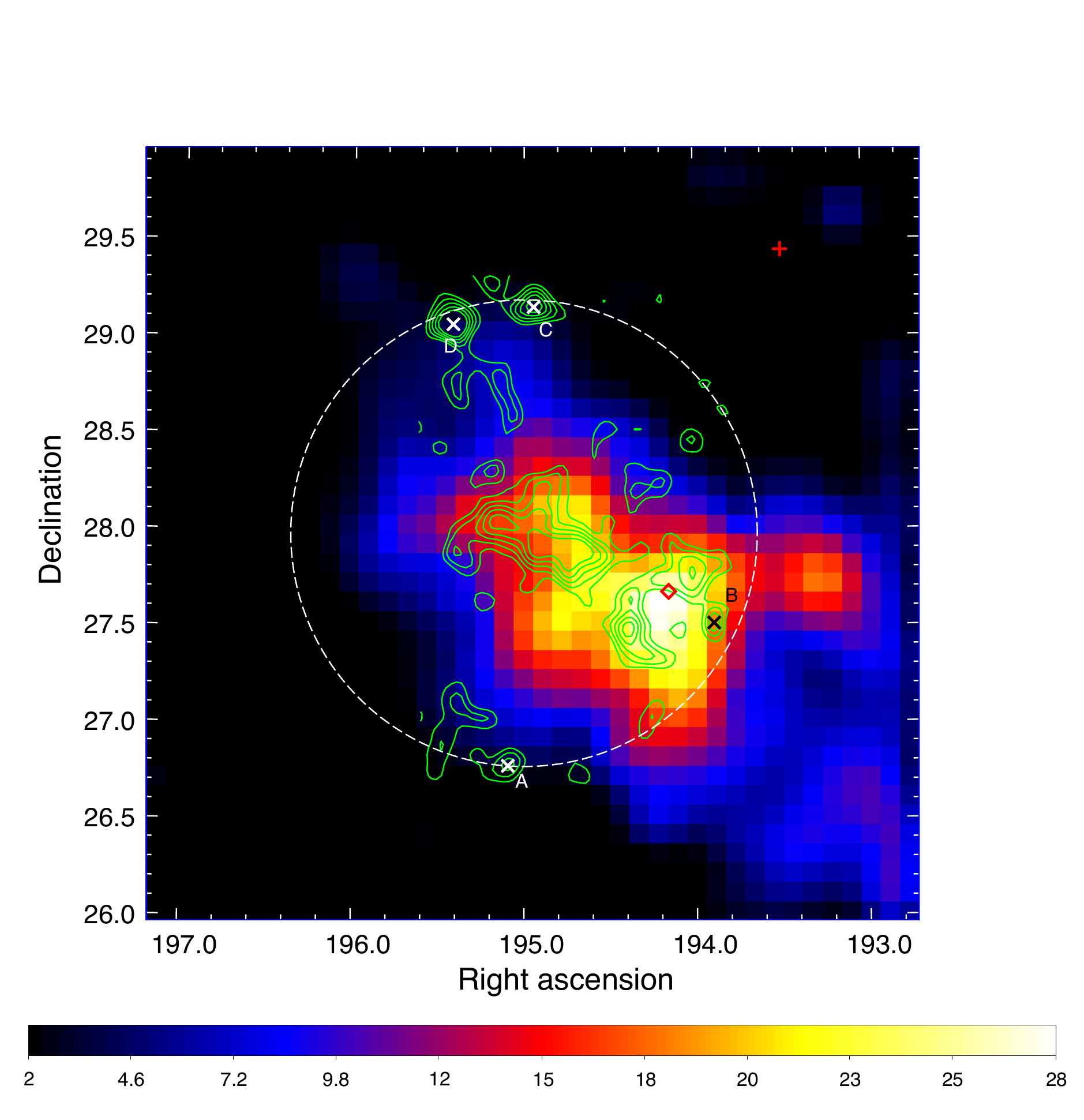}
\caption{The residual TS map of the Coma region after subtracting the background. The TS map is $4^\circ\times4^\circ$ with a pixel size of 0.02$^\circ$. The map has been smoothed with a 2 pixel gaussian kernel. The white dashed circle demonstrates the viral radius of Coma cluster. The green is the contour of the mass distribution around Coma region, which is derived by weak lensing observation \cite{Okabe14lensing}. Black and white crosses are 4 mass concentrations we will remove when generating spatial template for gamma-ray analysis, see text for detail. The red plus is a newly added point source, and the diamond is best fitted position assuming the residual is coming from a point source.}
\label{fig:tsmap}
\end{center}
\end{figure}

We use the script {\it make3FGLxml.py}\footnote{\url{http://fermi.gsfc.nasa.gov/ssc/data/analysis/user/}} to generate an initial background model, which include all 3FGL \cite{fermi15_3fgl} sources within $14^{\circ}$ around the Coma center as well as the latest galactic diffuse $\gamma$-ray emission model {\tt gll\_iem\_v06.fits} and isotropic emission template {\tt iso\_P8R2\_SOURCE\_V6\_v06.txt}\footnote{\url{http://fermi.gsfc.nasa.gov/ssc/data/access/lat/BackgroundModels.html}}. The spectral parameters of the 3FGL sources within $\rm 7^{\circ}$ ROI and the normalizations of the two diffuse backgrounds are set free in the likelihood fittings. Since the 3FGL is based on 4 years' Fermi-LAT data, while we are analyzing that of close to 9 years, to check whether there are new gamma-ray sources present in the current data set, we utilize the {\it gttsmap} tool to generate a residual TS map ({\it tsmap1}). Any excess in the {\it tsmap1} outside the virial radius of the Coma with a peak TS value $>25$ is treated as new point source, and is added into the background model with a power-law spectrum to model it. Their positions are determined using the {\it gtfindsrc} tool. Totally, three new point sources are identified within our ROI, all of which are also reported in \cite{fermi16_coma,xi17coma}. The fitted model containing above-mentioned 3FGL sources, diffuse gamma-ray components and 3 new point sources is our fiducial background model in the analysis. A TS map ({\it tsmap2}) based on the fiducial model is also generated to show the residual emission of the Coma after subtracting all background sources, which is shown in figure \ref{fig:tsmap}.

Beyond the fiducial background model, we add a Coma extended component with representative DM spatial and spectral models to test the significance of the emission. The spatial models applied in our work are listed in table \ref{tab1}, which will be described detailedly in the next section. The spectral model of DM annihilation follows the expression of
\begin{equation}
F(E_\gamma;\left<\sigma v\right>_f,m_\chi,f) = \frac{1}{4\pi} \frac{\left<\sigma v\right>_f}{2m_\chi^2} \frac{dN_f}{dE_\gamma}(E_\gamma) \times J,
\label{eg:flux}
\end{equation}
where $m_\chi$, $\left<\sigma v\right>_f$ and ${dN_f}/{dE_\gamma}$ are the DM mass, annihilation cross section and the differential gamma-ray production per annihilation for the annihilation final state $f$. In this work, we use PPP4DMID \cite{pppc4} to generate the DM spectra.

\section{Results}
In figure \ref{fig:tsmap} we demonstrate the residual TS map of the Coma region. The background model contains the components of 3FGL point sources (none is located in the $4^\circ\times4^\circ$ TS map region) and newly added point sources (one of which is denoted as red plus point in figure \ref{fig:tsmap}). The white circle shows the 1.2$^\circ$ virial radius of the Coma galaxy cluster, which is derived from the distance of $\sim105$ Mpc and $R_{200}$ of 2.2 Mpc. Clearly, in the TS map an excess with wide spreading and irregular shape appears within the virial radius, which doesn't look like a residual of point source.
For the hypothesis of this excess coming from DM, it's a natural expectation that the emission would trace the mass distribution in the Coma region, since the DM dominates the mass of galaxy cluster.
Thus we compare the residual TS map with the mass distribution derived through weak gravitational lensing observation \cite{Okabe14lensing} to show their correlation. The contour of the mass distribution is shown as green in figure \ref{fig:tsmap}. In the central part of the Coma region, the TS map residual is highly correlated to the mass distribution. At the edge of the viral circle, three structures with very high mass concentration (A, C and D in figure \ref{fig:tsmap}) don't show any evidence of gamma-ray emission. According to \cite{Okabe14lensing}, these three structures together with structure B are possibly overlapped with background systems. Considering their confusion and locating at the edge of the virial radius of the Coma, we will remove these 4 structures from the original weak lensing map of \cite{Okabe14lensing} when generating the template for likelihood analysis. we remove these structures with a simple way by truncating the map at the radius of 1$^\circ$.

\begin{table*}[t]
\caption{Likelihood fitting results of the Coma emission.}
\begin{ruledtabular}
\begin{tabular}{lccccc}
  Model & Flux [${\times10^{-9}\rm ph\,cm^{-2}s^{-1}}$] & Index & TS Value\footnote{TS value reported by {\it gtlike} tool.}  & DOF\footnote{Degree of freedom.}\\
\hline
Point Source\footnote{A single point source at the position of P3.}    & $0.962 \pm 0.220$  &  $-2.747 \pm 0.235$ & 35.7      &  4 \\
Disk            & $1.661 \pm 0.353$  &  $-2.679 \pm 0.241$ & 29.7      &  2 \\
NFW             & $0.971 \pm 0.248$  &  $-3.271 \pm 0.158$ & 20.8      &  2 \\
Lensing         & $1.402 \pm 0.276$  &  $-2.719 \pm 0.137$ & 36.3      &  2 \\
Disk+P3         & $1.167 \pm 0.483$  &  $-2.788 \pm 0.374$ & 14.0/18.1\footnote{The former/later one correspond to the TS value of the extended/P3 component, respectively.} &  2/4 \\
NFW+P3          & $0.643 \pm 0.271$  &  $-3.261 \pm 0.357$ & 9.4/24.1 &  2/4 \\
Lensing+P3      & $1.118 \pm 0.410$  &  $-2.885 \pm 0.169$ & 21.1/12.7 &  2/4 \\
\end{tabular}
\end{ruledtabular}
\label{tab1}
\end{table*}

To test whether the gamma-ray emission in Coma region is really correlated with the mass distribution or with other spatial models, we perform likelihood fittings to the Fermi-LAT data. The spatial models considered in our analyses are listed in table \ref{tab1} together with their corresponding fitting results. Here the {\it Lensing} model denotes using the mass distribution as spatial templates. The {\it Disk} model is a simple uniform disk with radius of $1.2^{\circ}$ (virial radius of Coma) centered on the Coma center. The presence of subhalos not only results in an enhancement of the J-factor but also flattens the surface brightness profile of the annihilation signal \cite{ando12_fornax,huang12cluster}. For simplicity, we use the disk to approximately model such a flattened gamma-ray emission profile. The $Disk$ model can also be treated as a generic model for any other spatial distribution.
The {\it NFW} model is a projection template that assume the DM in Coma following the Navarro-Frenk-White distribution \cite{navarro97nfw} with its $r_s$, $\rho_s$ determined by the cluster virial mass and the concentration-mass correlation \cite{sc14cm}. At last, for the {\it Point Source} model, we place a point source at the best fitted position of $\alpha_{\rm J2000}=194.15^\circ$, $\delta_{\rm J2000}=27.68^\circ$ (red diamond in figure \ref{fig:tsmap}). This point source is consistent with the P3 in X17, so we also use this name to denote it.

The significance of the target source can be quantified with the test statistic (TS), which is defined as $TS=-2{\rm ln}({L_{0}/L}$), where ${L}$ and $L_{0}$ are the maximum likelihood values for the models with and without target source, respectively.
The fourth column of table \ref{tab1} list the TS values of the tentative Coma emission for all the spatial models considered. Here we assume the spectrum is a power law and the fitted spectral index is listed in the table as well. Our fitting results display that the {\it Lensing} model gives the highest TS value of $\sim36$, only slightly higher than the PS one. Though the PS model gives comparable TS value, we want to emphasize that considering a higher degree of freedom (4 for {\it PS} vs. 2 for {\it Lensing}), its significance is even much lower than the latter. Thus a {\it Lensing} spatial distribution seems to be more favored. Moreover, if we add a P3 point source into the background model and fit the extended and P3 component simultaneously, the TS value of the {\it Lensing} is higher than the P3 one, implying the extended component can absorb more residual emission. This also supports the idea of the coma emission correlating with the mass distribution. Of course, the tentative signal is faint at the moment, it is not possible to distinguish these two models reliably. For the {\it NFW} and {\it Disk} spatial models, because they are both centered on the cluster center, while most of the residual emission appears in the south-west of the Coma region, these two models gives smaller TS value. Especially for the {\it NFW}, of which the expected emission is much more concentrated on the Coma center, the likelihood fitting gives the lowest TS value and very soft spectral index.
%We will also remind that the DOFs of these two models are smaller than the PS one, so the significances for them may be not lower than the later.

\begin{figure}[t]
\begin{center}
\includegraphics[width=0.95\columnwidth]{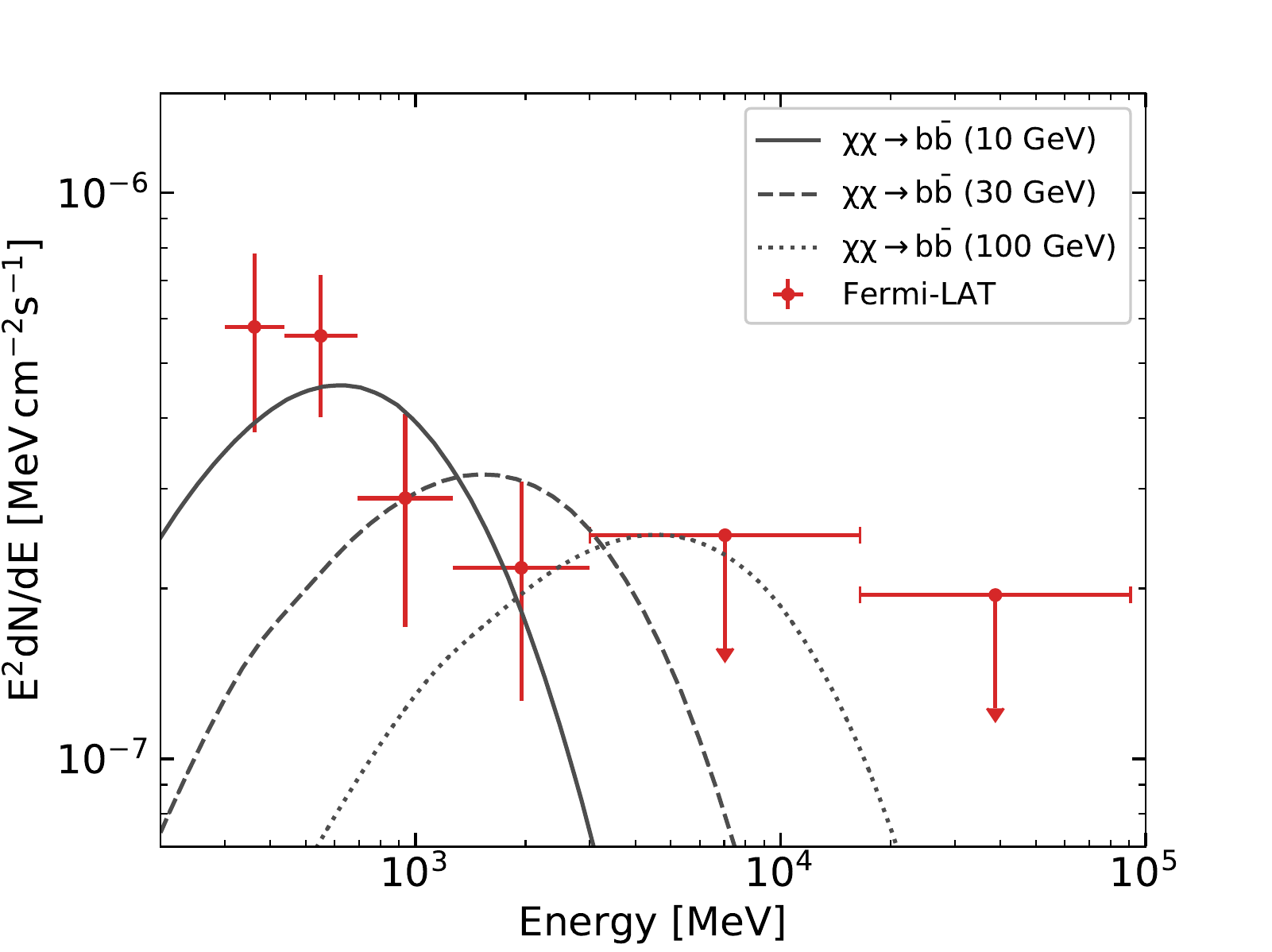}
\caption{The Fermi-LAT SED of the residual emission in the Coma region (red points). Upper limits are calculated when the TS value in that energy bin is small than 4. Black lines are some representative DM spectrum.}
\label{fig:sed}
\end{center}
\end{figure}

\begin{figure*}[htbp]
\begin{center}
\includegraphics[width=0.45\textwidth]{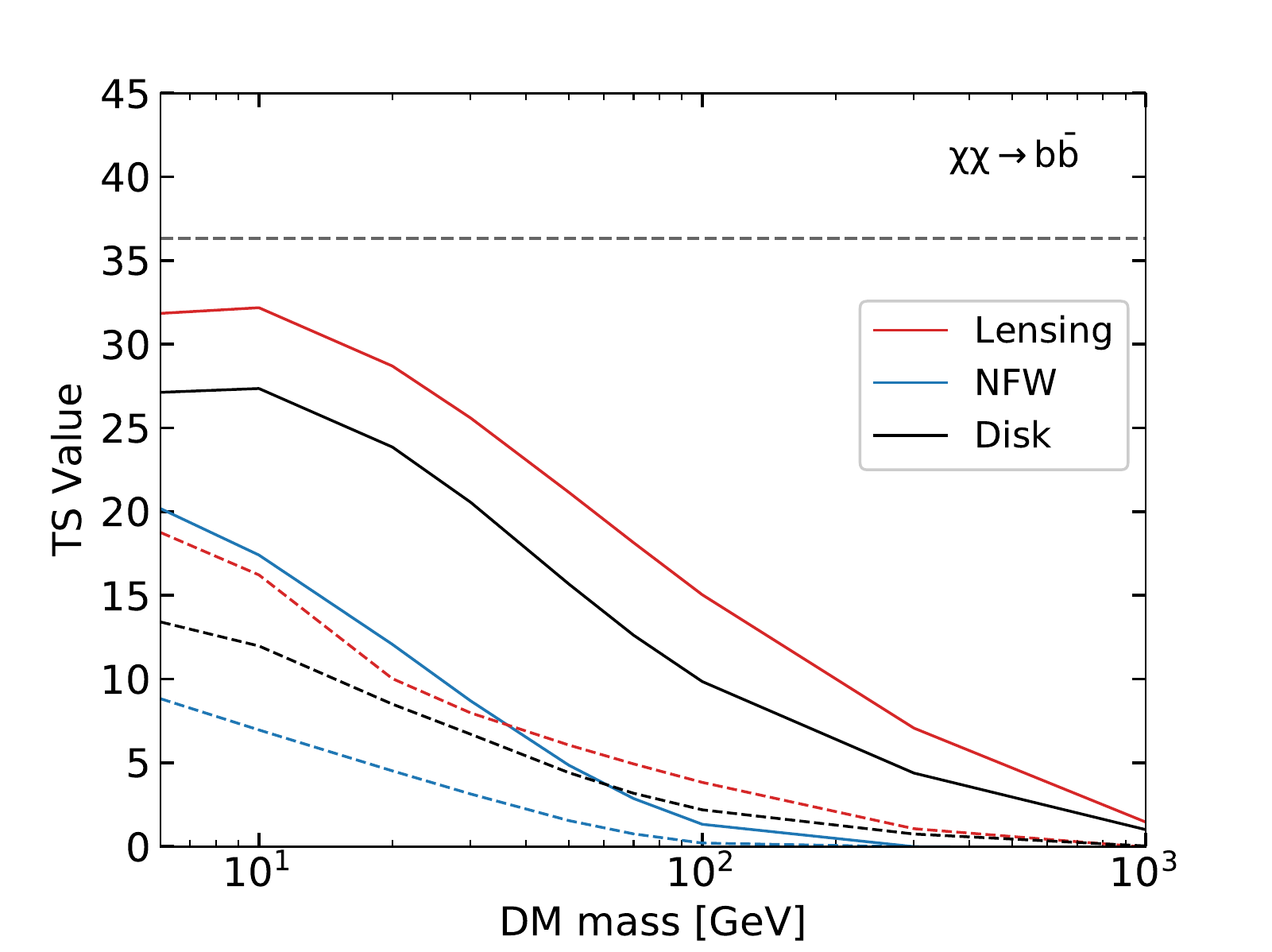}
\includegraphics[width=0.45\textwidth]{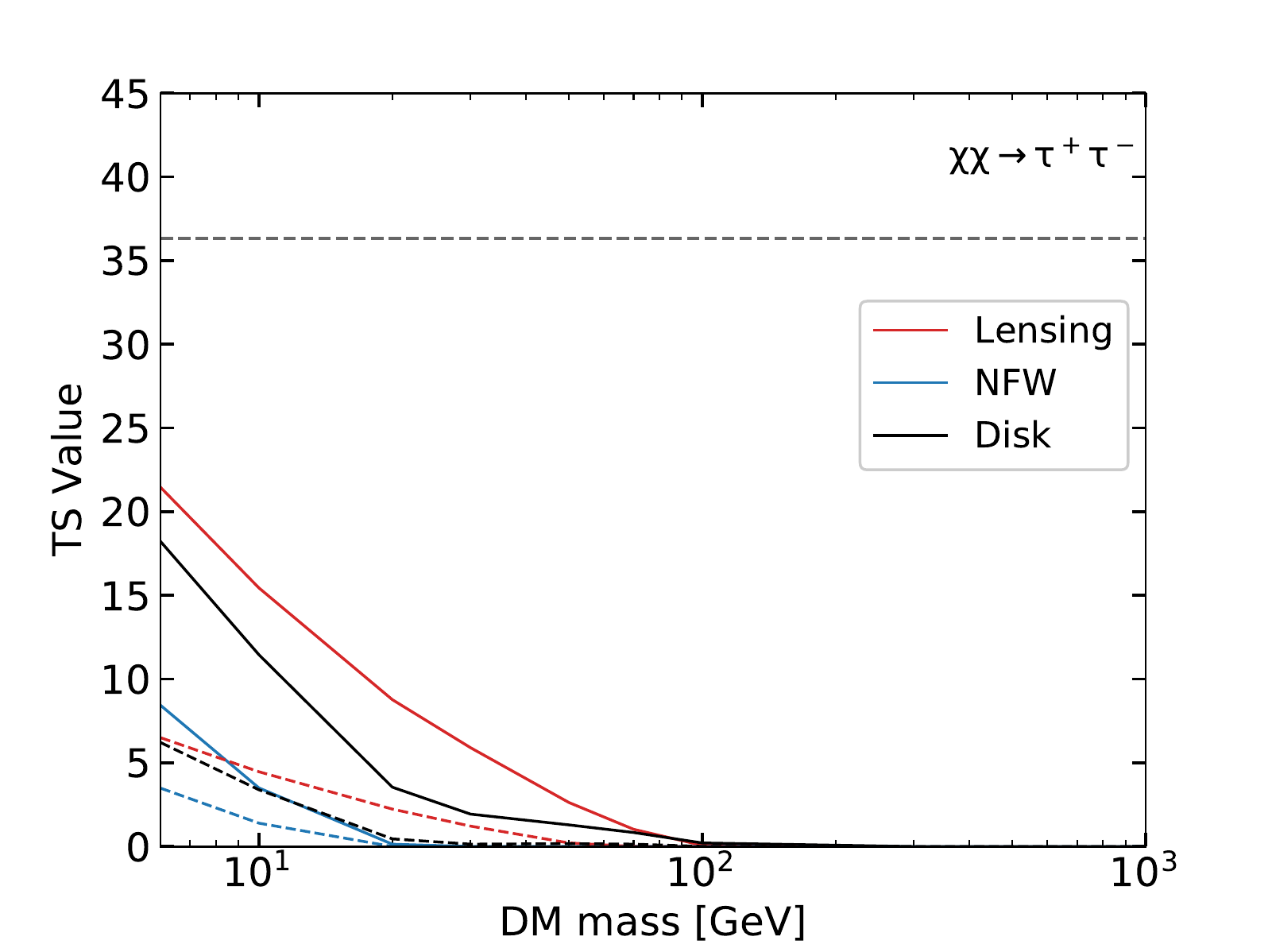}
\caption{The TS value of the putative extended Coma emission as a function of DM mass for models of $\chi\chi\rightarrow{b}\bar{b}$ and $\chi\chi\rightarrow{\tau^+\tau^-}$. Dashed lines are for the cases of P3 point source also being added into the background model. As a comparison, the TS value for the model with power law spectrum and weak lensing spatial distribution is also plotted (dashed horizontal line).}
\label{fig:tsvsmass}
\end{center}
\end{figure*}

Though the residual emission seems to be spatially correlated to the mass distribution (thus correlated to the DM distribution) according to above analysis, a DM origin requires that the spectrum matches the signal of DM annihilation as well. As listed in the table \ref{tab1}, the fittings using power law spectra give very soft spectral indexes, consistent with the results reported in X17.
Unfortunately, such soft spectra usually don't coincide with DM spectral models. To examine the spectrum more detailedly, we produce the spectral energy distribution (SED) of the extended component to compare it with some representative DM spectra. In figure \ref{fig:sed}, the SED of the Coma with {\it Lensing} spatial model is shown (red points). The expected DM spectra we present for comparison are the channel to $b$ quarks (black lines). From figure \ref{fig:sed}, the observed data seem to be a single power law with soft index rather than a bump which is the characteristics of DM annihilation spectrum. Therefore, models with large particle masses can not account for the fluxes in the low energy band, which are thus not favored. Alternatively, DM with a mass of 10 GeV in our demonstration seems to be able to produce spectrum match the observed data. The DM spectra expected by $\tau^+\tau^-$ final states for any particle masses are even harder, thus are also not favored.

We also perform likelihood fittings over a energy range of 300 MeV $-$ 500 GeV to test which DM spectral model gives the best fitting result. We fit the data for a grid of assumed DM particle masses and $b\bar{b}$, $\tau^+\tau^-$ annihilation final states. In figure \ref{fig:tsvsmass}, the TS value of the extended Coma component as a function of the DM mass is shown for different spatial models. The best fitted one corresponds to the spectrum of DM annihilating to $b\bar{b}$ and with a mass of 10 GeV, which is close to the minimum DM mass we consider. No matter what annihilation channel ($b\bar{b}$ or $\tau^+\tau^-$) and what spatial models, a general trend is lower mass DM model can fit the data better because of the power law behavior of the observed spectrum. In the aspect of spatial models, the {\it Lensing} spatial model results in the best fit due to the high correlation between the residual emission and the mass distribution given by weak lensing. Not surprisingly, adding a additional point source (P3) into the background models will largely decrease the significance of the extended components. However, not all residual emission can be absorbed by a single point source, the TS value of the extended components is still non-zero, though lower. Comparing with the fitting result using a power law spectrum and a $Lensing$ template (TS$\sim$36.3, horizontal dashed line, without P3 added), all the fittings adopting DM spectra give smaller TS values. This result indicates the observed data prefer other emission mechanism such as CR interaction within Coma\cite{xi17coma} to the DM hypothesis. Nevertheless, considering the difference of the maximum TS values between DM and PL models is small, there is still possibility of this result being due to statistical fluctuation that a DM origin is not absolutely excluded.

\begin{figure*}[t]
\begin{center}
\includegraphics[width=0.45\textwidth]{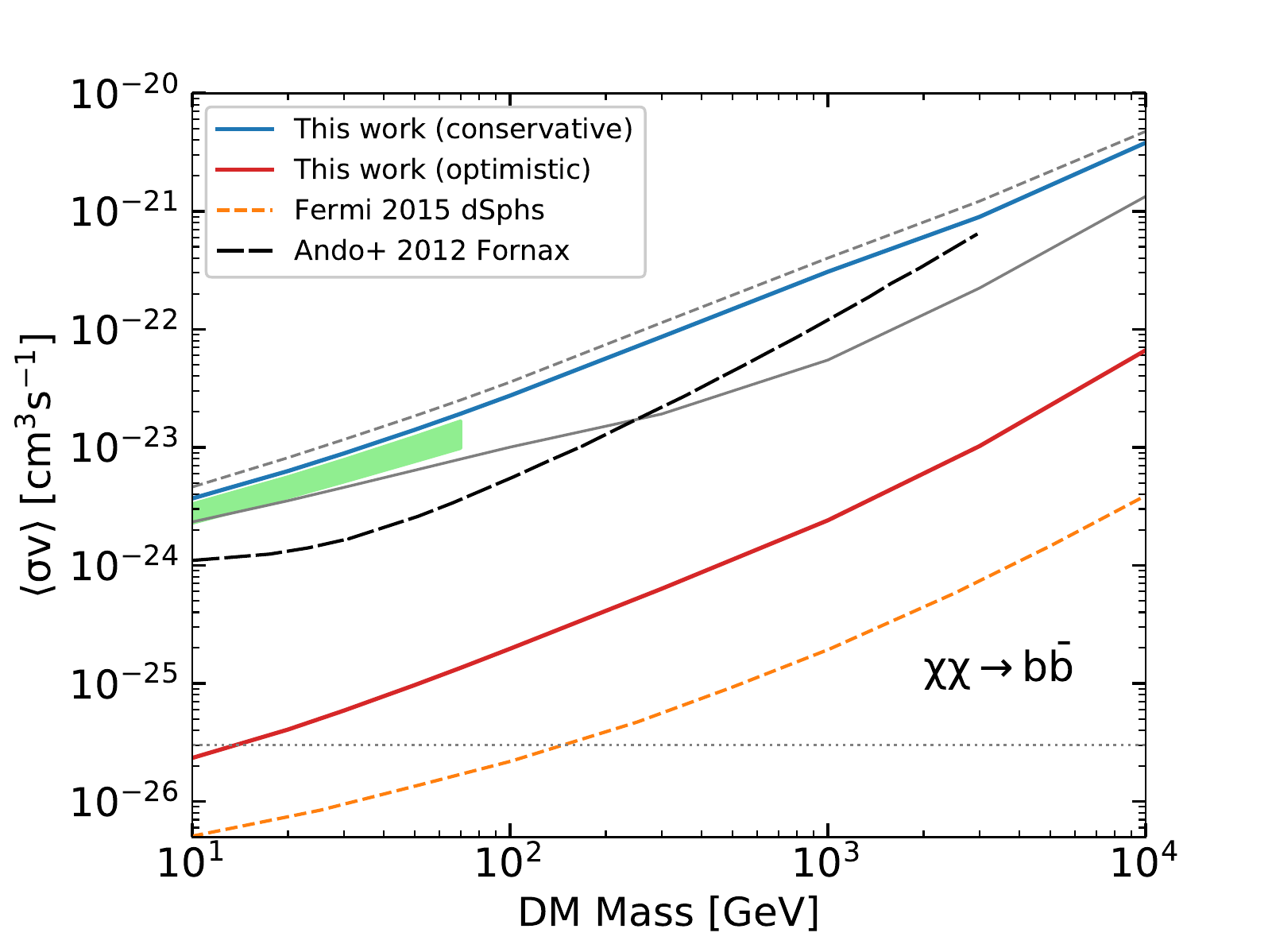}
\includegraphics[width=0.45\textwidth]{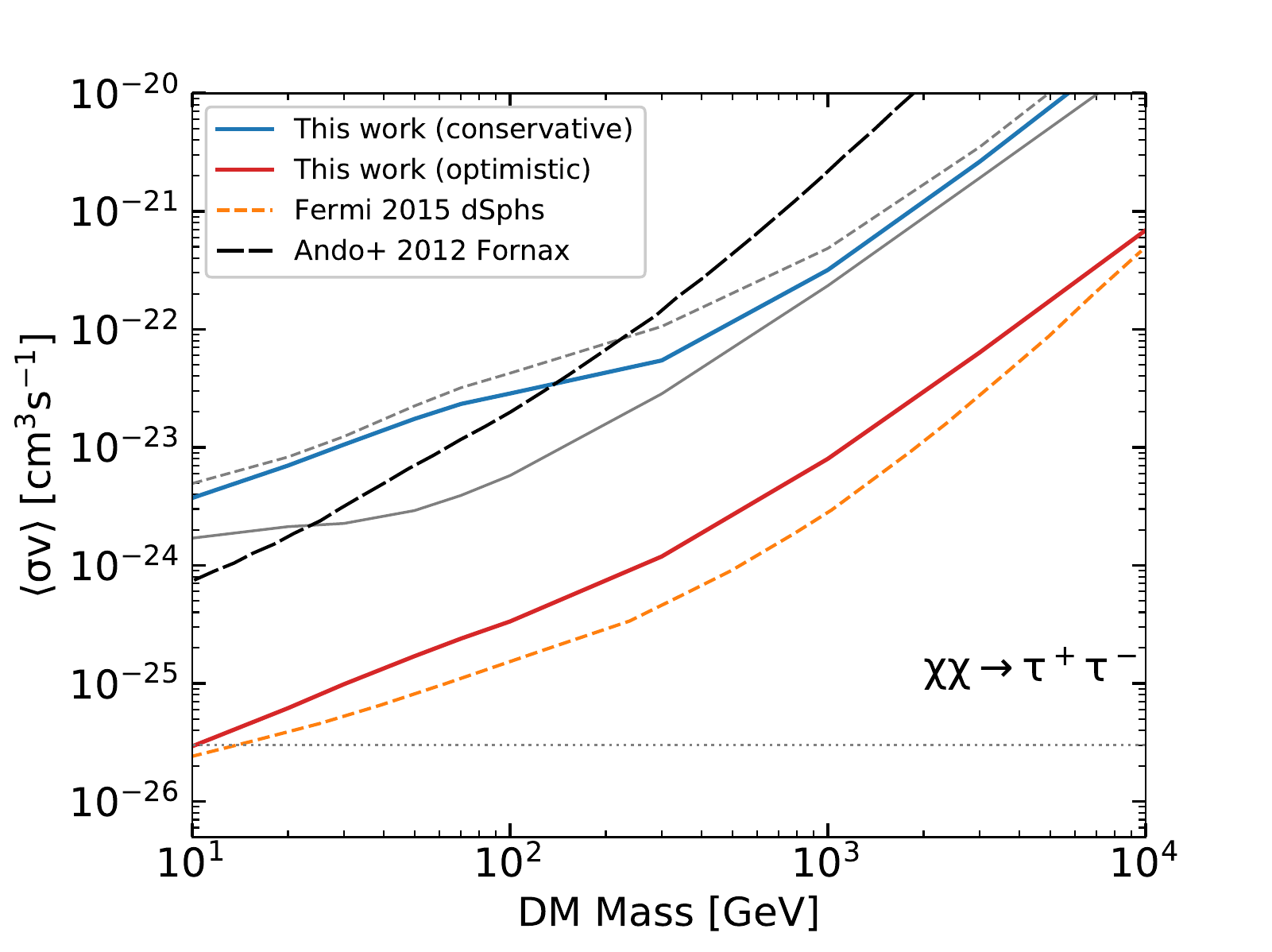}
\caption{Constraints on the DM annihilation cross section obtained in this work and comparison with those in the literatures. Red and blue are for the conservative and optimistic constraints respectively. When deriving the red/blue constraints, a {\it Lensing} spatial model is adopted. We also plot the results obtained using other spatial models; the dashed/solid gray lines are for {\it NFW}/{\it Disk}, respectively. The orange dashed lines are constraints by analyzing 25 milky way dwarf galaxy clusters by Fermi Collaboration \cite{fermi2015dsph6yr}, which is the current most stringent limits for $b\bar{b}$ and $\tau^+\tau^-$. The dotted line of cross section of $3\times10^{-26}\,{\rm cm^3s^{-1}}$ is roughly the thermal relic cross section. The green band in the left panel shows the mass range where the fittings result in TS value of the extended component greater than 16 and the corresponding $\left<\sigma{v}\right>$, the spreading in $y$ axis shows the 1 $\sigma$ uncertainty due to flux determination.}
\label{fig:sigmav_ul}
\end{center}
\end{figure*}

Disregarding the fact that DM model is somewhat not favored, we compare the DM properties could be derived assuming the residual emission is indeed coming from DM to the constraints set by other observations. Specifically, here we compare to the constraints set by Fermi-LAT observation of dSphs \cite{fermi2015dsph6yr}, since they are the most stringent ones for $b\bar{b}$ and $\tau^+\tau^-$ channel obtained by Fermi-LAT so far. In the left panel of figure \ref{fig:sigmav_ul}, the green band shows the DM annihilation cross section, $\left<\sigma{v}\right>$, derived from our analysis, {\it assuming the residual emission in the Coma region is indeed from DM}. We only consider the DM mass range where the fittings result in TS values greater than 16, and the spreading in $y$ direction is due to the uncertainties on the flux determination. As is shown, if the Coma emission was from DM, the required cross section is about 600 times higher than the current most stringent constraints (dashed orange line) if no boost factor is taken into account. To not conflict with the dSph constraints, a boost factor of $>$ 600 is needed for the Coma galaxy cluster. We note that such a boost factor is possible as predicted by some numerical $N$-body simulations \cite{Gao2012bf1e3}, while a modest BF predict by \cite{sc14cm} for Coma is $\sim35$. Although the conflict with the dSph constraints can be alleviated by including a relative high BF, we want to point out that the parameter space for the Coma emission can not survive the constraints by other galaxy clusters, such as Fornax (limits by Fornax without BF are shown as black dashed line), since Fornax also have a similar cluster mass (thus similar BF) to the Coma.

At last, because the DM models are not favored by the spectral analyses and limits of other GCls, we derive the constraints on the cross section of DM annihilating to $b$ quarks and $\tau$ leptons for the DM mass range from 6 GeV to 10 TeV.
To do this, firstly the 95\% confidence level (CL) flux upper limits at each particle mass are derived by scanning a series of {\it Prefactor} of the extended components and finding the {\it Prefactor} value that make the fitted log-likelihood vary by a factor of 1.35 compared to the maximum one.
The flux upper limits are then converted to the upper limits of $\left<\sigma{v}\right>$ according to the expression (\ref{eg:flux}). As for the $J$-factor, we use the value of $1.8{\times}10^{17}\,{\rm GeV^2\,cm^{-5}}$, which is derived assuming a NFW DM distribution and adopting the concentration-mass relation proposed by \cite{sc14cm}. Strictly, the $J$-factor derived here assuming a NFW profile should be valid for NFW model only. However considering the mass distribution by weak lensing measurements is complex and it's hard to derive its corresponding J-factor precisely, we also use NFW J-factor for this model. For disk model, which we use to describe the case of the surface brightness flattening due to the existence of subhalos \cite{ando12_fornax,huang12cluster}, the effect of subhalo will be taken into account by the boost factor. For the contribution from the smooth halo we will also adopt the NFW J-factor.

The figure \ref{fig:sigmav_ul} presents the upper limits based on the analysis of Coma cluster in this work. A {\it conservative} case is for assuming gamma-ray emission due to astrophysical processes such as cosmic ray interaction inside the cluster is absent. It is possible that the observed fluxes are entirely due to DM annihilation, thus we don't treat them as background when deriving the upper limits. We also ignore the effect of substructure for the conservative limits. In contrast, we consider a {\it optimistic} case of assuming all the Coma region extended emission and the possible P3 point source are real astrophysical sources, and subtract the corresponding contributions from the background first before deriving limits. To subtract the extended component, we adopt a radio template of the Westerbork Synthesis Telescope (WSRT) observation at 352 MHz \cite{Brown2010radioTemplate}, supposing the radio and gamma-ray emission coming from the same population of cosmic rays. The radio template used here is the same as that in X17. A boost factor of $\sim35$ is also taken into account for the optimistic case, but we don't consider the boost due to the DM contraction proposed by \cite{ando12_fornax}. In figure \ref{fig:sigmav_ul}, the blue and red curve are for conservative and optimistic constraints, respectively.
%The uncertainties propagated to the upper limits due to those on mass determination would be in the magnitude of 30\%-50\%, which is relatively negligible comparing the uncertainties from Boost factor. For better demonstration we didn't plot the uncertainties in the figure.

As some emissions appear within the Coma virial region, it is not surprising that the constraints on the dark matter $\left<\sigma{v}\right>$ obtained here are much worse than not only the one set by dSphs but also that by other galaxy cluster (i.e. Fornax) for the case of annihilating to $b\bar{b}$ and without a boost factor taken into account. However, for the channel to $\tau^+\tau^-$, our constraints are better in the high energy range comparing to Fornax limits.
Furthermore, if we adopt an {\it NFW} model, our analysis utilizing 9 years Fermi-LAT data set gives much stronger constraints (solid gray lines) than that of Fornax above 300/30 GeV for $b\bar{b}$/$\tau^+\tau^-$. This is because most of the emission for such spatial model is expected to come from very center of the Coma, where the observed emission is relatively weak.
Turn to the optimistic case, with a modest BF of $\sim35$, the limits according to the Coma observation can be close to current most stringent constraints for DM annihilating to $\tau^+\tau^-$.

\section{Summary}
In this work, we have analyzed 9 years Fermi-LAT observation of the Coma galaxy cluster to check whether the extended gamma-ray emission reported in \cite{xi17coma} can be attributed to a DM origin. By comparing the mass distribution inferred from weak lensing observations with the Fermi-LAT residual TS map within the Coma virial region, we found these two maps overlap with each other well.
However, further spectral analyses reveal a single power law spectrum with soft index can fit the data better than the spectra expected from DM annihilating to $b\bar{b}$ and $\tau^+\tau^-$. The former is more likely to come from astrophysical processes.
Further more, to account for the residual gamma-ray emission in the scenario of DM annihilation, a very large $\left<\sigma{v}\right>$ is needed which has been excluded by observations of dSphs even a boost factor of $\sim$35 is considered. Constraints by Fornax, which has a similar BF to the Coma cluster, can also excluded the parameter space that required.
All these facts indicate a DM interpretation to the observations is not favored, at least not all the emission come from DM particles.

Nevertheless, in this work we have confirmed the results reported by \cite{xi17coma} that there exist some extended residual emission with TS value $>25$ in the region around Coma, even with a higher lower energy bound of 300 MeV. We also show that for the channel to $\tau^+\tau^-$ the limits on $\left<\sigma{v}\right>$ based on our analysis of Coma cluster can be close to the current most stringent one considering a modest boost factor.

\begin{acknowledgments}
We thank Nobuhiro Okabe for sending us the weak lensing significance map and helpful introductions to the map. This work is supported in part by the National Key Research and Development
Program of China (No. 2016YFA0400200), the National Natural Science Foundation
of China (Nos. 11525313, 11722328, 11773075, U1738210, U1738136).
\end{acknowledgments}

\bibliographystyle{apsrev4-1-lyf}
\bibliography{coma}

\end{document}